\documentstyle[12pt]{article}
\catcode`\@=11
\begin{document} \openup6pt

\title{INFLATIONARY COSMOLOGIES IN AN ANISOTROPIC BRANE WORLD}

\author{Bikash Chandra Paul\thanks{Electronic mail : bcpaul@nbu.ernet.in}\\
	Physics Department, North Bengal University, \\
Siliguri, DIST. : Darjeeling, Pin : 734 430, West Bengal, INDIA \\
and \\
Inter-University Centre for Astronomy and Astrophysics \\
P.O. Box : 4, Ganeshkhind, Pune - 411 007, India}

\date{}

\maketitle

\vspace{0.5in}

\begin{abstract}

A new cosmological solution of the gravitational field equations in the generalized Randall-Sundrum model for
an anisotropic brane with Bianchi I geometry and with perfect fluid as matter sources is presented. The matter is described by a scalar field. The solution
admits inflationary era and at a later epoch the anisotropy of the universe
washes out. We obtain two classes of cosmological scenario, in the first case universe evolves from singularity and in the second case universe expands without singularity.
\end{abstract}

\vspace{0.2cm}

PACS number(s) : 04.50.+h, 98.80.Cq

\pagebreak

{\bf I. INTRODUCTION :}
  
Over the past couple of years much atention has been paid to studying cosmological models in higher dimensions motivated by developments in superstring and M-theory  [1,2].
In these theories, gravity is a higher dimensional theory which reduces effectively 4-dimensional at lower energies. Such higher dimensional theories opens up the possibility 
of solving the hierarchy problem in particle physics and make the string scale accessible to the future accelerators.  In such scenario our observed universe may be regarded as
 a domain wall  embedded in a higher dimensional spacetime. It is assumed that the fields of the standard model are confined to 3 + 1 dimensional hypersurface  (referred to as 3 brane)
embedded in  higher dimensional spacetime but the gravitational field may propagate through the bulk dimensions perpendicular to the brane. 

Recently a number of literature [3] appeared which explored the braneworlds with a homogeneous and isotropic cosmology. It was also found [4] that the Randall-Sundrum type
 brane world scenario can be extended to the dilatonic branes with Friedmann-Robertson-Walker (FRW) geometry.  In brane Cosmology the Friedmann equation gets modified by 
a term  quadratic  in density which enhances the possibility of getting more  inflation [5,6].  Maartens { \it et al.} [6]  studied chaotic inflation on the brane and demonstrated that the
 modified braneworld  Friedmann equation leads to a stronger condition for inflation. It is also noted by them that the brane effects ease the condition for slow-roll inflation 
for  a given potential.  The inflationary scenario in brane cosmology has the desired scale invariant Harrison-Zeldovich spectrum  for density perturbation in accordance
 with the recent  COBE data. 

The  behavior of an anisotropic Bianchi type-I brane world in the presence of a scalar field has been  explored by Maartens, Sahni and Saini [7] . The result obtained by them is
encouraging for  building cosmological models in brane-world.  It is demonstrated that a large anisotropy on the  brane does not prevent inflation, moreover, a large anisotropy
 enhances more damping into the scalar field equation of motion, resulting greater inflation.  Inflation at high energies proceeds at a higher rate than the corresponding rate
 in general relativity due to the presence of the quadratic term in energy density term in the modified Einstein equation.  The shear dynamics in Bianchi type I cosmological 
model on a brane with perfect fluid has been studied by Toporensky [8].  The result shows that for $1 <  \gamma  < 2$, the shear has a maximum during its transition from 
non-standard to standard cosmology i.e., when the matter energy density is comparable to the brane tension. Campos and Sopuerta [9] studied qualitatively both the Bianchi I and V cosmological solutions   in the Randall-Sundrum brane world scenario with matter on the brane obeying a barotropic  equation of state  $p = (\gamma  - 1) \rho $. It is found that anisotropic Bianchi I and V braneworlds always isotropize,  although there could be intermediate stages in which the anisotropy grows. In the models they found that near the bigbang, the anisotropy dominates for  $ \gamma  \leq 1 $, which is a different result from that in the general relativistic situation where it happens for $ \gamma  \leq 2 $.  Frolov [10] has carried out geometrical construction of the Randall-Sundrum braneworld, without the assumption of spatial isotropy but by means of a homogeneous and anisotropic Kasner type solution of the Einstein-Ads equation in the bulk. Santos {\it et al.} [11] 
obtained condition to be satisfied by the brane matter and bulk metric so that a homogeneous and anisotropic brane evolves asymptotically to a de Sitter space-time in the presence of a positive cosmological constant. 

In this paper I present an exact new solution of the gravitational field equation in the braneworld model for an anisotropic Bianchi type-I geometry for a conformally flat bulk (with vanishing Weyl tensor). The inclusion of the quadratic terms in the energy-momentum tensor of the perfect cosmological fluids leads to major changes in the early dynamics of the anisotropic universe, as compared to the standard general theory of Relativity (GTR) case.

The paper is organized as follows : The field equations on the brane are written down in sec. II. In sec. III we present the field equation for anisotropic 
brane. Evolution of the scalar field in the braneworld and exact solution of the field equation are given in sec IV. Finally, a brief discussion is given in sec. V.

{\bf II. BRANE GEOMETRY AND FIELD EQUATION:}

The Einstein's field equation in the  five dimensional (bulk) framework can be written in the following form : 
\begin{equation}
G_{AB}^{(5)} = \kappa_{(5)}^{2} \left[ - g_{AB}^{(5)} \Lambda_{(5)} + 
 T_{AB}^{(5)} \right] 
\end{equation}
with $T_{AB}^{(5)} = \delta (y) [ - \lambda g_{AB} + T_{AB} ]$. We use to represent the upper case 
Latin letters for coordinate indices in the bulk spacetime ($A, B, ... = 0, ..., 4$)  and small case latin letters for three space ($i, j = 1, 2, 3 $) whereas the Greek letters for the coordinate indices in the four dimensional spacetime where matter is confined ($\mu, \nu, ... = 0, ..., 3$). We will use physical units in which $c = 1$. In equation (1) $\kappa_{(5)} $  represents the five dimensional gravitational coupling constant; $g_{AB}^{(5)}$, $
G_{AB}^{(5)} $ and  $\Lambda_{(5)} $ are the metric, Einstein tensor and the cosmological constant of the bulk space-time respectively, $T_{AB}$ is the matter energy momentum tensor. Here $\kappa^{2}_{(5)} = \frac{8 \pi}{M_{P}^{3}} $ where $M_{P} = 1.2 \times 10^{19} $ GeV. A natural  choice of coordinates is $x^{A} = ( x^{\mu}, y ) $ where $x^{\mu} = (t, x^{i})$ are space-time coordinates on the brane. The space-like hypersurface $x^{4} = y = 0$ gives the brane world and $g_{AB}$ is its induced metric, $\lambda $ is the tension of the brane which is assumed to be positive in order to recover conventional general
theory of gravity (GTR) on the brane. The bulk cosmological constant 
$\Lambda_{(5)} $ is negative and is the only five dimensional source of the gravitational field. 

The field equations induced on the brane are derived using geometric approach 
[12] leading to new terms which carry bulk effects on to the brane. The modified Einstein's equations of General Relativity on the brane is
\begin{equation}
G_{\mu \nu}^{(5)} = - \Lambda g_{\mu \nu} + \kappa^{2} T_{\mu \nu}
   + \kappa_{(5)}^{4} S_{\mu \nu} - E_{\mu \nu}^{(5)} .
\end{equation}
The four dimensional gravitational constant $k$ and the cosmological constant $\Lambda$ are given by
\[
k^{2}  = \frac{1}{6} \lambda \kappa_{(5)}^{2},
\]
\begin{equation}
\Lambda = \frac{|\Lambda_{5}|}{2}  \left[  \left( \frac{\lambda}{\lambda_{c}} \right)^{2} - 1 \right] 
\end{equation}
respectively, where $\lambda_{c}$ is the critical brane tension which is given by 
\begin{equation}
\lambda_{c} = 6 \frac{|\Lambda_{5}|}{\kappa_{5}^{2}} .
\end{equation}
In the above one finds that the extra dimensional corrections to the Einstein equations on the brane are of two types : 

$\bullet $
$S_{\mu \nu} $ are corrections quadratic in the matter variables which is
\begin{equation}
S_{\mu \nu} =  \frac{1}{12} T T_{\mu \nu} - \frac{1}{4}  T_{\mu \alpha} T^{\alpha}_{\nu} + \frac{1}{24} g_{\mu \nu} \left[ 3 T_{\alpha \beta} T_{\alpha \beta} - (T^{\alpha}_{\alpha})^{2} \right] .
\end{equation}
where $T = T^{\alpha}_{\alpha}$, $S_{\mu \nu}$ is significant at high energies 
i.e., $\rho > \lambda$, 

$\bullet$
 $E_{\mu \nu}^{5} $ are corrections due to the non-local effects from the free gravitational field in the bulk, which  comes into play via the projection 
${\bf E}^{(5)}_{AB} =  C^{(5)}_{ACBD} n^{C} n^{D}$ where $n^{A}$ is normal to the surface ($n^{A} n_{A} = 1$), which is symmetric and traceless and without components orthogonal to the brane, so ${\bf E}_{AB} n^{B} = 0$ and 
${\bf E}_{AB} \rightarrow  E_{\mu \nu} g^{\mu}_{A} g^{\nu}_{B} $
as $y \rightarrow 0$.

All the local and the non-local  bulk corrections may be consolidated into an effective total energy-momentum tensor. The modified Einstein equations take the standard Einstein form :
\begin{equation}
G_{\mu \nu} = - \Lambda g_{\mu \nu} + \kappa^{2} T_{\mu \nu}^{total}
\end{equation}
where $ T_{\mu \nu}^{total} = T_{\mu \nu} + \frac{6}{\lambda}
   S_{\mu \nu} - \frac{1}{\kappa^{2}} E_{\mu \nu} $ .
The effective total energy density, pressure, anisotropic stress and energy flux are [13] 
\[
\rho^{total} = \rho \left(  1 + \frac{\rho}{2 \lambda} \right) + \frac {6 {\bf U}}{\kappa^{4} \lambda },
\]
\[
p^{total} = p + \frac{\rho}{2 \lambda} (\rho + 2 p) +  + \frac {2 {\bf U}}{\kappa^{4} \lambda },
\]
\[
\pi^{total} = \frac {6 }{\kappa^{4} \lambda } {\bf P_{\mu \nu}}, 
\]
\begin{equation}
q^{total}_{\mu} = \frac {6 }{\kappa^{4} \lambda } {\bf Q_{\mu \nu}}
\end{equation}
where $ {\bf U}$ is the effective non-local energy density on the brane, $
 {\bf P_{\mu \nu}}$ is the effective non-local anisotropic stress and 
$ {\bf Q_{\mu \nu}}$ is the effective non-local energy flux on the brane that 
carries Coulomb and gravito-magnetic effects from the free gravitational field in the bulk. The brane energy-momentum tensor and the overall effective energy momentum tensor are both conserved separately. We are interested here in the particular  case of a Bianchi-I brane geometry which is a simplest generalization of a Friedmann-Robertson-Walker (FRW) brane geometry.

{\bf III. FIELD EQUATIONS FOR BIANCHI I BRANE :}

Consider an anisotropic brane-world model with flat three space of Bianchi 
type-I spaces. The Bianchi-I brane has the induced metric :
\begin{equation}
ds^{2} =  -  dt^{2} +  \sum_{i=1}^{3} R_{i}^{2} (dx^{i})^{2}
\end{equation}
and is covariantly characterized by
\[
D_{\mu} f = 0, \; A_{\mu} = \omega_{\mu} = 0, \; {\bf Q_{\mu}} = 0 ,\; 
R_{\mu \nu}^{*} = 0
\]
where $ D_{\mu} $ is the projected covariant spatial derivative, $f$ is any physically defined scalar, $A_{\mu } $ is the four-acceleration, $\omega_{\mu} $  is the vorticity and $R_{\mu \nu}^{*} = 0$ is the Ricci tensor of the 3-surface orthogonal to $u^{\mu}$.

The conservation equations reduce to 
\[
\dot{\rho} + \Theta (\rho + p) = 0
\]
\[
\dot{{\bf U}} +\frac{4}{3} \Theta {\bf U} + \sigma^{\mu \nu} P_{\mu \nu} = 0
\]
\begin{equation}
D^{\nu} P_{\mu \nu} = 0
\end{equation}
where a dot denotes $u^{\nu} \bigtriangledown_{\nu}$, $\Theta$ is the volume expansion rate, $\sigma_{\mu \nu} $ is the shear. The directional Hubble parameters are defined by
\[
H_{i} = \frac{\dot{R_{i}}}{R_{i}}
\]
and the mean expansion factor $a = \left( R_{1} R_{2} R_{3} \right)^{\frac{1}{3}} $. One gets the expansion rate 
\begin{equation}
\Theta = 3 H = 3 \frac{\dot{a}}{a} = \sum_{i=1}^{3} H_{i} ,
\end{equation}
the average anisotropy expansion is given by
\begin{equation}
A = \frac{1}{3} \sum_{i = 1}^{3} \left (\frac{\triangle H_{i}}{H}  \right)^{2}.
\end{equation}
For a conformally flat bulk geometry one can set ${\bf U}  = 0$ at the cost of  $\sigma^{\mu \nu} P_{\mu \nu} = 0$.

For the metric (8)  one gets
\begin{equation}
\sigma^{\mu \nu} \sigma_{\mu \nu} = \sum_{i=1}^{3} (H_{i} - H)^{2} = \frac{6 \Sigma^{2}}{a^{6}} , \; \; \;  \dot{\Sigma } = 0,
\end{equation}
and the  anisotropy parameter becomes
\begin{equation}
A = \frac{\Sigma^{2}}{H^{2} V^{2}}
\end{equation}
where $V = a^{3}$.
We now use equation  
(12) to obtain generalized Friedmann equation for the Bianchi-I brane (with
$\Lambda = {\bf U} = 0$) which is  given by 
\begin{equation}
H^{2} = \frac{8 \pi G}{3}  \rho \left( 
1 + \frac{\rho}{2 \lambda} \right)  + \frac{\Sigma^{2}}{a^{6}}
\end{equation}
we replaced here $\kappa^{2} $ by $8 \pi G$. One recovers the field equation corresponding to the brane FRW world when $\Sigma = 0$ [14].

{\bf IV. COSMOLOGICAL SOLUTION :}

We now describe the matter content in the universe with a homogeneous scalar field $\phi = \phi (t)$ on the brane. The energy momentum tensor is
\begin{equation}
T_{\mu \nu} = \rho u_{\mu} u_{\nu} + p g_{\mu \nu} 
\end{equation}
with $\rho = \frac{1}{2} \dot{\phi}^{2} + V(\phi)$ and $p = \frac{1}{2} \dot{\phi}^{2} - V(\phi)$, where  $\frac{1}{2} \dot{\phi}^{2}$ is the kinetic energy and $ V(\phi)$ is the potential energy. The evolution equation for $\phi (t)$ is
\begin{equation}
\ddot{\phi} + 3 H \dot{\phi} + \frac{dV (\phi)}{d \phi} = 0.
\end{equation}
We now look for an exact cosmological solution. For this we take a constant potential  $ V(\phi) = V_{o} = Constant $. In this case the equation (16) can be integrated  to yield
\begin{equation}
\dot{\phi} =  \frac{C}{a^{3}}
\end{equation}
where $C$ is an integration constant. Equation (14) now can be written as
\begin{equation}
\left( \frac{\dot{a}}{a} \right)^{2} = \beta^{2} \left(1 + \frac{\alpha}{a^{6}} \right) \left(\eta_{1} + \frac{\eta \alpha}{a^{6}} \right) + \frac{\Sigma^{2}}{a^{6}}
\end{equation}
where we use the symbols for simplification $\beta = \sqrt{\frac{8 \pi G V_{o}}{3}}$, $\alpha = \frac{C^{2}}{2 V_{o}}$,
$\eta = \frac{V_{o}}{2 \lambda} $ and  $\eta_{1} = 1 + \eta$. 

To obtain solution of the differential equation (18) we substitute $y = ln \;
a(t)$ and obtain  
\begin{equation}
\frac{dy}{dt} = \pm \beta \sqrt{ \eta_{1} +  \eta_{2}  e^{-6y} +
 \eta_{3}  e^{-12 y} }
\end{equation}
where $ \eta_{2} = \alpha (\eta +  \eta_{1})  + 
\frac{\Sigma^{2}}{\beta ^{2}}$,  $\eta_{3} = \eta \alpha^{2} $. On integrating equation (19) yields :
\begin{equation}
a^{6}(t) =  \frac{B}{4 \eta_{1}} e^{6 \sqrt{\eta_{1}} \beta t}  
   + 
\frac{  \eta_{2}^{2} - 4  \eta_{1} \eta_{3}}{4   \eta_{1} B}
  e^{- 6 \sqrt{\eta_{1}} \beta t } 
- \frac{\eta_{2}}{2 \eta_{1}}.
\end{equation}
where $B$ is an integration constant. The above solution is  new in anisotropic brane cosmology. The intersting aspect of the solution is that it depends on the brane parameter $\eta$ which is effectively related to the brane tension $\lambda$. 
At late times  the solution asymptotically approaches de Sitter expansion. Thus it would be interesting to study the solution both at 
high and at low energy limits to compare the behavior of the early universe in the brane-world and in GTR.
Different values of the integration constant $B$ which is determined by the boundary condition, lead to different classes of cosmologies which we discuss next.

{\bf Model I :  Expansion from a singularity :}

The cosmological solution given below 
\begin{equation}
a(t) =  \left( \frac{1}{4} \frac{(\eta_{2} + 2 
\sqrt{\eta_{1} \eta_{3} })}{
 \eta_{1} } e^{6 \sqrt{\eta_{1}} \beta t}     + 
\frac{  \eta_{2}^{2} - 4  \eta_{1} \eta_{3}}{4 (  \eta_{2} + 2 \sqrt{ \eta_{1 } 
 \eta_{3}}) \eta_{1}}  e^{- 6 \sqrt{\eta_{1}} \beta t } 
- \frac{\eta_{2}}{2 \eta_{1}} \right)^{\frac{1}{6}}
\end{equation}
corresponds to a universe evolving from a singularity which is obtained for $ B = \eta_{2} + 2 
\sqrt{\eta_{1} \eta_{3} }$. 

At late times $t \rightarrow \infty$, the solution asymptotically approaches de Sitter solution. Thus it would be interesting to study the solution both at high and low energy limit to compare the behavior of the early universe in brane-world and in GTR.
 In the low energy limit i.e., as $\frac{V_{o}}{\lambda} \rightarrow 0$, the symbols used above in (19) now take the values :
$ \eta \rightarrow 0 $,  $ \eta_{1} = 1$, $  \eta_{2} = \alpha + 
\frac{ \Sigma^{2}}{\beta^{2}}$ and $  \eta_{3} \rightarrow 0 $
leading to a simple solution 
\begin{equation}
a(t) = \left( \alpha + \frac{ \Sigma^{2}}{\beta ^{2}} 
\right)^{1/6} Sinh^{1/3 } (3 \beta t).
\end{equation}
The evolution of the scalar field is given by
\begin{equation}
\phi = \phi_{o} + \phi_{1} \; ln \left( tanh \frac{3 \beta}{2} \right) t,
\end{equation}
where $\phi_{o}$ is an integration constant and $\phi_{1} = C/\left( \alpha + \frac{ \Sigma^{2}}{\beta ^{2}}
\right)^{1/2} $. The anisotropy of the universe  decreases as
\begin{equation}
A = \frac{\Sigma^{2}}{(\alpha \beta^{2} + \Sigma^{2} ) Cosh^{2} (3 \beta t)}.
\end{equation}

In the high energy limit, when 
(i) $\Sigma = 0$ and $\alpha \neq 0$, corresponds to the field equation in the FRW-brane and (ii) $\Sigma \neq 0$ and $\alpha \neq 0$
corresponds to Bianchi-I anisotropic brane. It admits another case with a constant scalar field $\phi = \phi_{o}$ which corresponds to $\alpha = 0$ and $\Sigma \neq 0$.  We obtain solutions in these cases which will be discussed below.
The solutions obtained here are  interesting in the framework of braneworld. These solutions also admit two more classes of solutions
in the low energy limit  as a special case : (i)
when $\Sigma = 0$ and $\alpha \neq 0$  the solution corresponds to that obtained by Faraoni [15], (ii) $\Sigma \neq 0$ and $\alpha \neq 0$ corresponds to the solution discussed by Gron [16] in the context of isotropization of an anisotropic Bianchi I universe in GTR. We discuss the solutions next.

In the high energy limit, $\eta \rightarrow \infty$, the brane effects are most important which leads to an exponential expansion
\begin{equation}
a(t) = \left[ \left( \alpha +  \frac{\Sigma^{2}}{4 \eta \beta^{2}} \right) e^{6 \sqrt{\eta} \beta t}  -  \left( \alpha + \frac{\Sigma^{2}}{2 \eta \beta^{2}} \right) \right]^{1/6}.
\end{equation}
If $ \frac{\Sigma^{2}}{4 \beta^{2} \alpha } << \eta $, one gets a simple solution
\begin{equation}
a(t) = \alpha^{1/6} \left[ e^{6 \sqrt{\eta} \beta t} - 1 \right]^{1/6}.
\end{equation}
Thus the solution  admits inflationary scenario for a finite value of the anisotropy determined by the brane parameter. In the FRW braneworld model, Sami [17] reported similar solution. 

The scalar field evolves as
\begin{equation}
\phi (t) = \phi_{o} + \frac{C}{3 \beta \sqrt{\eta \alpha} } tan^{-1} \sqrt{ e^{6 \sqrt{\eta} \beta t} - 1 }.
\end{equation}
 Thus in the low energy limit  the scale factor evolves as  $a(t) \sim t^{1/3}$ whereas that in the high energy limit $a(t)  \sim t^{1/6}$. This reveals the phenomena that the singularity approaches more slowly in brane world model either in the presence of anisotropy or in the absence of anisotropy ($\Sigma = 0$, brane-FRW) than in the Friedmann model.

{\bf Model II :  Expansion without singularity :}

The cosmological solution (20) permits a universe evolving without singularity when one chooses  integration constant $B = 2
\eta_{1} a_{o}^{6} + \eta_{2} + 2 \sqrt{\eta_{1}^{2 } a_{o}^{12} + 
 \eta_{1} \eta_{2} a_{o}^{6} +  \eta_{1} \eta_{3}} $
 with $a (t = 0) = a_{o}$. Now   consider a simple case 
 $ \eta_{2}^{2} = 4  \eta_{1} \eta_{3}$, in this case the average scale factor evolves as
\begin{equation}
a(t) = \left[ \left( a_{o}^{6}  +  \frac{ \eta_{2}}{2 \eta_{1}} \right)
e^{6 \sqrt{\eta_{1}} \beta t } -  \frac{ \eta_{2}}{2 \eta_{1}} \right]^{1/6}. 
\end{equation}
In the low energy limit,
as $ \frac{V_{o}}{\lambda} \rightarrow 0$, the symbols now take the values :
\[
\eta \rightarrow 0, \; \; \eta_{1} = 1, \; \;  \eta_{2} = \alpha + \frac{ \Sigma^{2}}{\beta^{2}}, \; \;  \eta_{3} \rightarrow 0.
\]
the solution (28) now becomes
\begin{equation}
a = \left[ (a_{o}^{6} + \frac{1}{2} (\alpha + \Sigma^{2}/\beta^{2})) e^{6 \beta t} - \frac{1}{2} (\alpha + \Sigma^{2}/\beta^{2}) \right]^{1/6}.
\end{equation} 
The scalar field in this scenario evolves as
\begin{equation}
\phi(t) = \phi_{o} + \frac{C}{3(\alpha + \Sigma^{2}/\beta^{2}) \beta} ln 
\left( \frac{ \left( a_{o}^{6} + 1/2 (\alpha + \Sigma^{2}/\beta^{2}) \right)
  e^{6 \beta t} - 1/2 ( \alpha + 
\Sigma^{2}/\beta^{2})}{\left( a_{o}^{6} + 1/2 ( \alpha + \Sigma^{2}/\beta^{2}) \right)
  e^{6 \beta t}}
 \right).
\end{equation}
 However in the high energy limit $\eta \rightarrow \infty$ one gets de Sitter universe in brane world which evolves as
\begin{equation}
a(t) = \left[ (a_{o}^{6} + \alpha) e^{6 \sqrt{\eta} \beta t} - \alpha \right]^{1/6}.
\end{equation}
The rate of expansion in this regime is greater than that in the low energy limit.

{\bf V. Discussions :}

A  class of cosmological solutions in the Bianchi-I brane world model are presented. The braneworld scenario described here admit two types of cosmological models, in one case universe expands from singularity and in the other case universe expands without singularity. Both the models correspond to a constant scalar field cosmological constant. The cosmological solutions are studied both in the high energy and in the low energy limits. In the low energy limit some of the solutions are identified with the exact solutions that are previously obtained in GTR. We compare the brane solution with the solution in GTR, in the case of a universe evolving from a singularity. It  is found that near the origin the singularity in Brane world model approaches more slowly $a(t) \sim
t^{1/6}$ than that obtained in GTR model $a(t) \sim t^{1/3}$.  It is  found that an inflationary universe may be obtained in the presence of large anisotropy which subsequently washes out in the inflationary era. Thus at a large time an isotropic and homogeneous universe emerges.

\vspace{0.5in}

{\large \it Acknowledgement :}
The work is supported by  the Minor Research Project grant by the University Grants Commission, New Delhi and North Bengal University.
I would like to  thank the Inter-University Centre
for Astronomy and Astrophysics (IUCAA) Pune for providing a facility where this work was carried out. 
\pagebreak

\end{document}